# Intra- and extra-axonal axial diffusivities in the white matter: which one is faster?

Nicolas Kunz, Analina R. da Silva and Ileana O. Jelescu[*]

*Centre d'Imagerie Biomédicale, École Polytechnique Fédérale de Lausanne, Lausanne, Switzerland*

A two-compartment model of diffusion in white matter, which accounts for intra- and extra-axonal spaces, is associated with two plausible mathematical scenarios: either the intra-axonal axial diffusivity $D_{a,\|}$ is higher than the extra-axonal $D_{e,\|}$ (Branch 1), or the opposite, i.e. $D_{a,\|} < D_{e,\|}$ (Branch 2). This duality calls for an independent validation of compartment axial diffusivities, to determine which of the two cases holds. The aim of the present study was to use an intracerebroventricular injection of a gadolinium-based contrast agent to selectively reduce the extracellular water signal in the rat brain, and compare diffusion metrics in the genu of the corpus callosum before and after gadolinium infusion. The diffusion metrics considered were diffusion and kurtosis tensor metrics, as well as compartment-specific estimates of the WMTI-Watson two-compartment model. A strong decrease in genu $T_1$ and $T_2$ relaxation times post-Gd was observed ($p < 0.001$), as well as an increase of 48% in radial kurtosis ($p < 0.05$), which implies that the relative fraction of extracellular water signal was selectively decreased. This was further supported by a significant increase in intra-axonal water fraction as estimated from the two-compartment model, for both branches ($p < 0.01$ for Branch 1, $p < 0.05$ for Branch 2). However, pre-Gd estimates of axon dispersion in Branch 1 agreed better with literature than those of Branch 2. Furthermore, comparison of post-Gd changes in diffusivity and dispersion between data and simulations further supported Branch 1 as the biologically plausible solution, i.e. $D_{a,\|} > D_{e,\|}$. This result is fully consistent with other recent measurements of compartment axial diffusivities that used entirely different approaches, such as diffusion tensor encoding.

**Keywords**: diffusion MRI; modeling; white matter; duality; compartment diffusivities.

## 1. Introduction

The diffusion MRI signal is sensitive to the micron-scale displacement of water molecules in tissue and can thus provide valuable information about the underlying microstructure. However, because water is ubiquitous, modeling is required to infer compartment-specific diffusion metrics. Modeling involves assuming a simplified geometry of the tissue under consideration and fitting the analytical expression of the diffusion signal in such an environment to the measured data.

Since the aim of biophysical models is to provide a more specific characterization of microstructure than signal representation approaches such as diffusion tensor imaging (DTI) [1] and diffusion kurtosis imaging (DKI) [2], biophysical modeling has drawn great attention and research efforts in the past years [3-11].

However, modeling also implies making simplifying assumptions which, if incorrect, can heavily bias the estimation and impact the interpretation of the result [12-14]. For example, white matter, which is the subject of the current study, is typically described by two or three compartments (**Figure 1**): the intra-axonal space is modeled as a collection of infinitely long cylinders with a given orientation distribution function, the extra-axonal space is assumed to behave as a Gaussian anisotropic medium, and the cerebrospinal fluid (CSF) contribution – if accounted for – is modeled as a Gaussian isotropic compartment with fixed free diffusivity $D_{iso}$ = 3 μm²/ms *in vivo*. The parameter estimation for the apparently "simple" two-compartment model has been shown to present two substantial issues: there are two distinct mathematical and biologically-plausible solutions to the system, and each solution is surrounded by a gentle-sloped optimization landscape, whereby noise can displace the minimum by a significant amount from the ground truth [15, 16]. In qualitative terms, the two solutions of the two-compartment model can be described as one where $D_{a,\parallel} < D_{e,\parallel}$ and axon dispersion is limited (i.e. $c_2 \equiv \langle(\cos\psi)^2\rangle$ is close to 1), and another where $D_{a,\parallel} > D_{e,\parallel}$ and axon dispersion is more pronounced.

[*] ileana.jelescu@epfl.ch

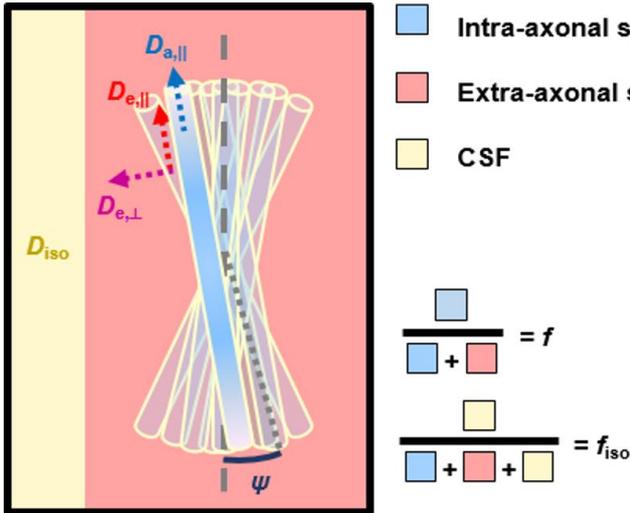

**Figure 1.** Schematic diagram of a typical three-compartment white matter model. Neurite sub-bundles have a given orientation distribution of angles ψ about the main bundle axis (vertical axis in the figure). The local diffusivities within each sub-bundle are denoted as $D_{a,\parallel}$ for the intra-axonal compartment ($D_{a,\perp}=0$), and $D_{e,\parallel}$ and $D_{e,\perp}$ for the extra-axonal compartment. $f$ is the fraction of intra-axonal water. The CSF compartment can also be accounted for as a third compartment, with relative water fraction $f_{iso}$ and diffusivity $D_{iso}$ = 3 μm²/ms in vivo.

The unveiled degeneracy calls for an independent validation of compartment axial diffusivities, to establish which of the two solutions mimics better the biological reality.

This task is arguably more challenging than validating axonal water fractions or even orientation dispersion, because alternative methods to NMR for measuring the self-diffusion coefficient of water are not available, and NMR-based water measurements include signals from all compartments. Notwithstanding, numerous efforts have been undertaken towards achieving compartment-specific diffusivity measurements. Methods for achieving compartment-selectivity were initially designed to explain the dramatic decrease in mean diffusivity (MD) during stroke [17-19]. Recently, the focus of such research has shifted precisely towards finding the correct solution to the two-compartment white matter model [20-25].

In this work, we build on the idea initially explored by Silva *et al*. to suppress the extracellular signal by injecting a gadolinium (Gd)-based contrast agent in the lateral ventricles of the rat brain and thus measure diffusion weighted signals that stem mostly from the intracellular space [19]. In a diffusion experiment, the amount of MR signal stemming from a given compartment $C$ is weighted not only by the physical amount of water $V_C$ in that compartment, but also by the compartment $T_2$, i.e. $S_C \propto V_C \cdot e^{-\frac{TE}{T_{2,C}}}$. The extra-cellular gadolinium is thus expected to preferentially shorten the $T_2$ of the extra-cellular compartment (though the $T_2$ of the intra-cellular space will likely also be somewhat reduced) and thereby decrease the contribution of the extra-cellular space to the overall measured signal. In their work, Silva *et al.* performed measurements in a subcortical area in the rat gray matter using a maximum of three orthogonal directions to estimate the trace of the diffusion tensor. They reported no significant change in mean diffusivity after extracellular signal suppression with gadolinium and concluded that intra- and extra-cellular diffusivities were similar. Here, we focus on the rat corpus callosum, and estimate changes in diffusivity and kurtosis in the axial and radial directions (relatively to the main fiber orientation) separately. Furthermore, we estimate specific changes in the metrics of a WMTI-Watson two-compartment model of diffusion [20]. The goal of this work was to determine whether the changes resulting from the attenuation of the extracellular signal are compatible with the $D_{a,\|} > D_{e,\|}$ or $D_{a,\|} < D_{e,\|}$ scenario.

## 2. Methods
### 2.1. Animal preparation

This study was approved by the Service for Veterinary Affairs of the canton of Vaud. Nine adult Sprague-Dawley rats (270 ± 13 g, 6 males) underwent two MRI sessions, two days apart. The first session was dedicated to baseline measurements of diffusion and relaxometry ("pre-Gd") and the second session, consisting in the same measurements, started one hour after an intracerebroventricular perfusion of gadolinium ("post-Gd"). Rats were sacrificed at the end of the post-Gd MRI session.

### 2.2. Intracerebroventricular perfusion

The methods were similar to those described by [18, 19]. The rat was anesthetized with isoflurane (4% for induction and 2% for maintenance) and positioned in a stereotaxic frame (Kopf Model 900). The head was shaved and the skull was exposed with a midline incision. Two holes were burred into the skull using a dental drill (~1.5 mm Ø), giving access to the two lateral ventricles (stereotaxic coordinates: ±1.4 mm lateral, 0.9 mm posterior and 3.5 mm deep relative to the bregma). A volume of 20 uL (10 uL per ventricle) of 0.25 M gadobutrol (Gadovist, Bayer Schering Pharma) was perfused continuously over two hours (rate: 5 uL/hour in each ventricle) using Hamilton syringes, a double-syringe pump (TSE System) and in-house catheters with 30G needles. At the end of the perfusion, the catheters were removed and the skin was sutured with silk sutures 3/0. The rat was given one hour to recover from surgery before starting the MRI acquisitions – isoflurane anesthesia was maintained throughout.

### 2.3. MRI acquisition

The protocol was the same for both scanning sessions, with the exception of relaxometry parameters that were adjusted to accommodate different ranges of $T_1$ and $T_2$ values pre- and post-Gd (Table 1). The anesthetized rat was transferred and fixed in a homemade MRI cradle equipped with a fixation system (bite bar and ear bars). Anesthesia was maintained at 1.5 – 2% isoflurane in an air-oxygen mixture (50% oxygen / 50% medical air) throughout the experiment, to ensure a breathing rate of around 60 bpm, which was continuously monitored using a respiration pillow placed under the animal's thorax. Body temperature was also continuously monitored using a rectal thermometer and maintained around (38 ± 0.5) °C using a warm water circulation system.

All MRI experiments were performed on a 14-T Varian system (Abingdon, UK) equipped with 400 mT/m gradients. An in-house built quadrature surface coil was used for transmission and reception.

|  | Pre-Gd | | | Post-Gd | | |
| --- | --- | --- | --- | --- | --- | --- |
|  | $T_1$ short TE | $T_1$ long TE | $T_2$ | $T_1$ short TE | $T_1$ long TE | $T_2$ |
| TR (ms) | 15,000 | | | 7,000 | | |
| TE (ms) | 2.8 | 30 | 2.8 – 150 (18 values) | 2.8 | 30 | 2.8 – 120 (14 values) |
| TI (ms) | 10 – 10,000 (19 values) | | - | 6 – 6,000 (19 values) | | - |
| NA | 8 | | | | | |

**Table 1.** STEAM sequence parameters for the measurement of $T_1$ and $T_2$ relaxation times in the genu.

Sagittal gradient-echo images were acquired to assess the success of the perfusion and the extent of gadolinium transport throughout the brain (TE/TR = 4/100 ms ; matrix: 128 x 128 ; 0.18 x 0.18 mm² in-plane resolution; 7 0.8-mm slices; flip angle: 20°).

Shimming was first performed in a voxel encompassing the genu of the corpus callosum (1.5 x 2 x 2 mm³), using FASTMAP and FASTESTMAP [26, 27]. The water linewidth was 14 ± 2 Hz in the pre-Gd sessions, while in the post-Gd sessions it varied around 41 ± 13 Hz at the time of relaxometry measurements (1.5 hours after perfusion). A smaller voxel fitted to the genu (1 x 1.5 x 1.5 mm³) was used for single-voxel $T_1$ and $T_2$ relaxometry using a STEAM sequence. $T_1$ was measured for two different echo times (TE = 2.8 ms and TE = 30 ms), the longer echo aiming to achieve extracellular water signal suppression in the post-Gd session, similar to the long echo time in the diffusion acquisition (see below). All sequence parameters are collected in **Table 1**.

For imaging, the field homogeneity was adjusted in a 5 x 6 x 10 mm³ region of interest encompassing the corpus callosum. The water linewidth was 27 ± 3 Hz in the pre-Gd sessions, and 70 ± 18 Hz in the post-Gd session, at the time of the diffusion data acquisition (3 hours after gadolinium perfusion). Diffusion data (4 b=0; b=1 and b=2 ms/μm² with 20 directions each) were acquired using an in-house semi-adiabatic spin-echo EPI sequence [28] with following parameters: TE/TR = 48/2000 ms, matrix: 128x64, FOV: 23x17 mm², 4 shots, 5 sagittal 0.8-mm slices, δ/Δ = 4/20 ms, NR = 6, TA = 47 min.

### 2.4. Data processing and analysis

Water spectra were quantified using the AMARES tool in jMRUI [29] and resulting signals were fit to the monoexponential decay (for $T_2$) and recovery (for $T_1$) models in Matlab.

Diffusion images – amounting to 44 *q*-space points x 6 repetitions – were first denoised using random matrix theory [30] before averaging over repetitions. The noisemap was further used for Rician bias correction $S_f = \sqrt{|S_i^2 - \sigma^2|}$ (σ: noise level, $S_i$: denoised signal, $S_f$: corrected signal) [31], prior to fitting the diffusion and kurtosis tensors using a weighted linear least squares algorithm [32]. A mask for the genu was manually drawn on the mid-sagittal slice (where the orientations of the axons within the bundle are likely to be most coherent) and mean values for typical DKI metrics (fractional anisotropy, mean/axial/radial diffusivities and kurtoses) were extracted in this ROI, for each rat and each session (pre- and post-Gd).

The WMTI-Watson two-compartment model was used to estimate compartment-specific metrics pre- and post-Gd [20]. This model assumes a Watson distribution of axon orientations, whereby the ODF is fully characterized by a single parameter. Model parameters $f$, $D_{a,\|}$, $D_{e,\|}$, $D_{e,\perp}$ and $\langle(\cos\psi)^2\rangle \equiv c_2$ can be expressed analytically as a function of main tensor metrics (axial and radial diffusivities, axial and radial kurtosis), as derived in [16, 20]. The equations are reproduced as Supplementary Data S1. This model has two possible solutions [15, 16, 20], both of which were reported.

Paired *t*-tests were used to compare pre- and post-Gd relaxation times and diffusion metrics in the genu.

Finally, the angle between the principle eigenvector of the diffusion tensor and the **B**$_0$ field was calculated in order to provide an assessment of the influence of Gd-driven susceptibility gradients on diffusion estimates.

### 2.5. Simulations

Median pre-Gd estimates of the WMTI-Watson model were used as ground truths for simulating diffusion signals in Matlab [20]. The post-Gd ground truth was identical to the pre-Gd with the exception of an increased intra-axonal water fraction, consistent with experimental estimates. The simulations matched the experiment in terms of diffusion protocol, SNR and Rician noise. The denoising procedure applied to the data resulted in a noise reduction of $\sqrt{P(1/M + 1/N)}$ (with *P*: number of significant components in the Marchenko-Pastur distribution, *M*: number of diffusion measurements and *N*: number of voxels in sliding window) [30] which was accounted for in the simulations. One thousand noise realizations were generated in each case.

Simulated signals were processed identically to the experimental signals in terms of tensor estimation and WMTI-Watson model estimation. In the experimental data, values are averaged over the genu ROI in each animal before statistical analysis. To reproduce the reduction in noise resulting from this ROI averaging, simulation results were bundled in groups of 20 and averaged, yielding 50 simulated "genu ROI" estimates.

The simulations were designed to capture the effects of an increase in intra-axonal fraction on all the model parameters, given realistic ground truths (for both Branch 1 and Branch 2) and noise levels. The comparison of trends between experiments and simulations can give further insight into the effects of post-Gd susceptibility gradients, which are present in the data but not in simulations, and also evaluate which of the two branches is realistic.

## 3. Results
### 3.1. Gadolinium perfusion

To assess the efficacy of Gd administration, gradient-echo images were acquired, which demonstrated accumulation of gadolinium in both lateral ventricles, while the concentration levels in the genu remained low enough to preserve image quality (**Figure 2**), also in the diffusion-weighted spin-echo EPI images (**Supplementary Figure 1**). The delays between icv perfusion, relaxometry and diffusion acquisitions were similar for all rats, and $T_2$ and diffusion measurements were performed within 66 ± 23 min of each other.

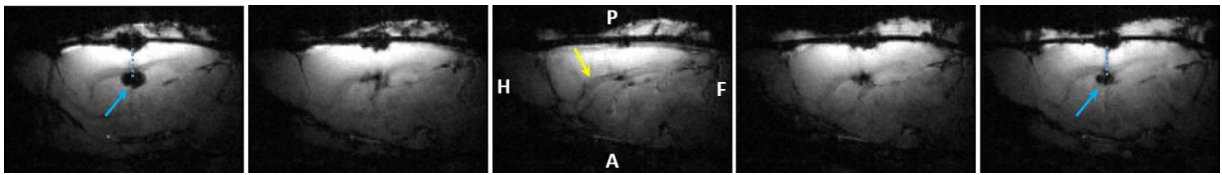

**Figure 2.** Contiguous sagittal gradient echo images, showing contrast agent accumulation in both lateral ventricles (blue arrows) yet acceptable image quality around the genu of the corpus callosum (yellow arrow). The path of the catheters is illustrated by the dotted blue lines: the surgical procedure is not expected to damage the genu and to allow gadolinium to enter the intra-axonal space.

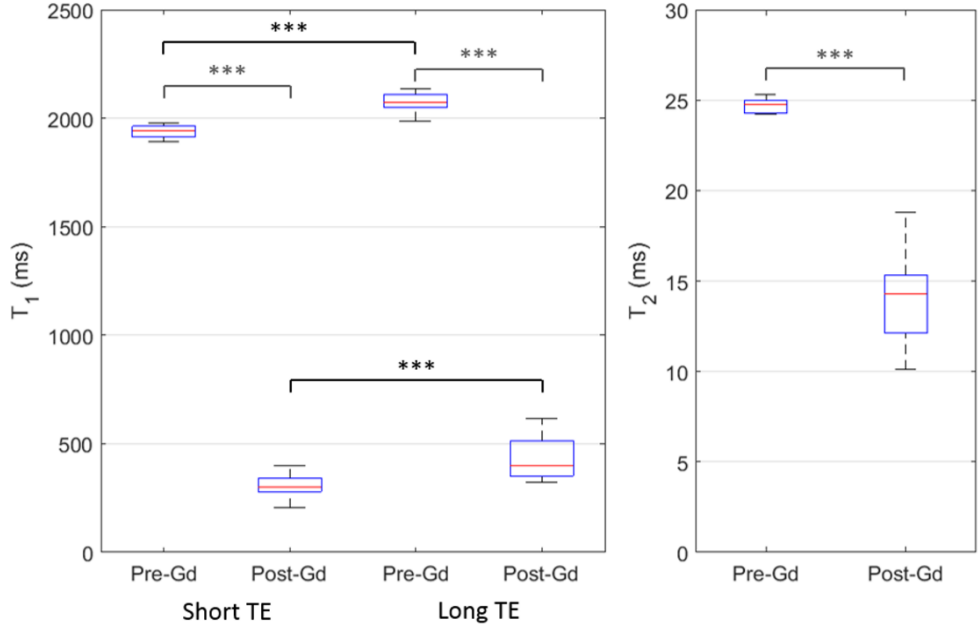

**Figure 3**. $T_1$ (left) and $T_2$ (right) relaxation times measured in the genu, before and after gadolinium perfusion. Relaxation times were significantly shortened by the presence of gadolinium in the extracellular space. The $T_1$ was significantly longer in the long-TE (30 ms) vs short-TE (2.8 ms) measurement. Pre-Gd, the difference can be attributed to suppression of myelin water with the longer TE; post-Gd, we hypothesize the long TE suppressed the extracellular contribution. It is noteworthy that, in the long TE measurement, the post-Gd $T_1$ was nonetheless much shorter than the pre-Gd $T_1$, indicating exchange between compartments and imperfect compartment selectivity. *Red line: median; Box edges: 25th and 75th percentiles; Whiskers: extreme datapoints*. ***: $p < 0.001$.

The $T_1$ and $T_2$ relaxation times in the genu were significantly shortened post- vs pre-Gd ( and **Figure 3**), which confirmed the accumulation of contrast agent. The variability in relaxation time measurements was much larger post-Gd – depending on the efficacy of each individual perfusion. The $T_1$ was significantly longer for the long vs the short TE acquisition, both pre- and post-Gd.

### 3.2. Diffusion and kurtosis tensors

Of the extracted tensor parameters, the only statistically significant difference between pre- and post-Gd conditions was the increase in radial kurtosis (RK) by 48% ($p < 0.05$) (**Figure 4**).

Given the very high reproducibility of $T_2$ measurements between rats in the pre-Gd condition, the measured post-Gd $T_2$ was used as a proxy for Gd concentration in the extra-cellular space, to assess direct correlations between contrast agent accumulation and tensor metrics. Linear correlations between $T_2$ and diffusion metrics were significant only for RK (Pearson's $\rho = -0.66$; $p < 0.01$) and mean kurtosis (MK) (Pearson's $\rho = -0.59$; $p < 0.05$)

|  | $T_1$ (short TE) (ms) | $T_1$ (long TE) (ms) | $T_2$ (ms) |
|---|---|---|---|
| Pre-Gd | 1938 ± 29 | 2073 ± 48 | 24.6 ± 0.4 |
| Post-Gd | 306 ± 63 | 430 ± 99 | 14.1 ± 2.7 |

**Table 2**. $T_1$ and $T_2$ relaxation times measured in the genu, before and after gadolinium perfusion.

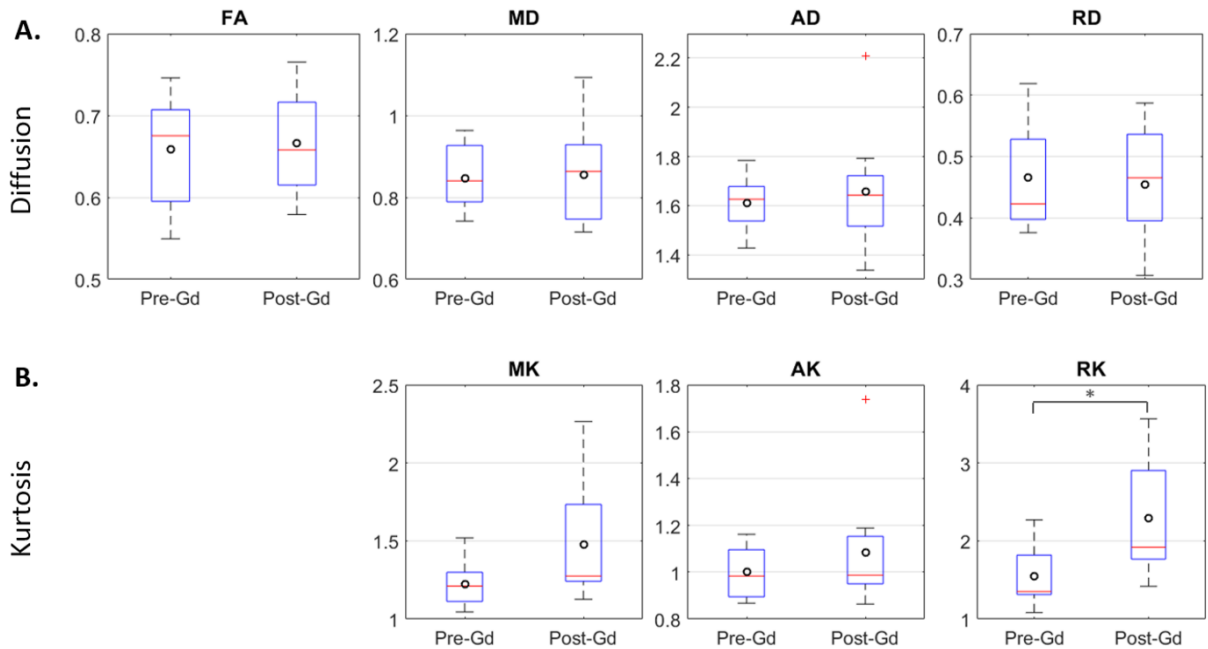

**Figure 4**. Diffusion tensor (A) and kurtosis tensor (B) metrics in the genu, before and after gadolinium perfusion. The radial kurtosis increased significantly post-Gd, which was consistent with an increased intra-axonal relative signal fraction. None of the diffusion metrics were significantly altered by the perfusion. *Red line: median; box edges: 25th and 75th percentiles; whiskers: extreme datapoints; red cross: outliers; black circle: mean. *: p < 0.05*. FA = fractional anisotropy; MD/AD/RD = mean/axial/radial diffusivity (in µm²/ms); MK/AK/RK = mean/axial/radial kurtosis.

(**Supplementary Figure 2**), which is consistent with previous group differences.

### 3.3. Diffusion tensor vs B₀ orientation

The corpus callosum fibers run left-right and the expected angle between the fibers' main orientation and the magnetic field $\mathbf{B}_0$ is 90°. This angle is fairly immune to most common sources of experimental variability in rat head positioning, which would be rotations about the x-axis (chin tilt due to variations in bite-piece positioning) or z-axis ("ear-to-shoulder" tilt due to misaligned ear bars).

In the pre-Gd session, the principal orientation of the diffusion tensors in the genu formed an angle largely comprised between 84° and 104° with the main field (**Figure 5**). The mode of the distribution was 94° and the median angle was (93 ± 4)° across rats, which could point to a slight but systematic misalignment of our stereotaxic system with the main field. In the post-Gd session, the distribution of angles was broader, with most orientations comprised between 76° and 112°, and a median angle of (96 ± 7)° across rats. This is consistent with previous reports of unaccounted gradients affecting the measured main orientation of the diffusion tensor [33].

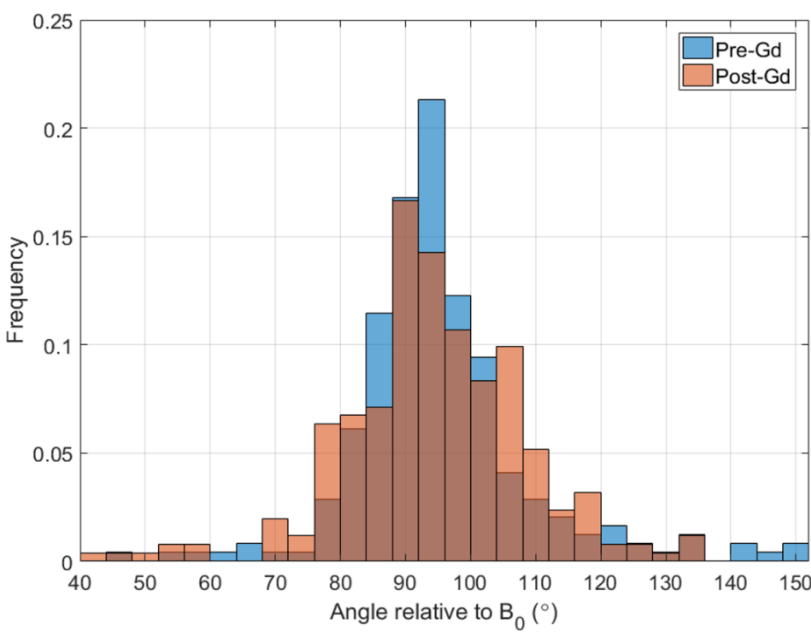

**Figure 5**. Distribution of angles between the principle direction of the diffusion tensor and the main field $\mathbf{B}_0$, across all genu voxels in all rats. The post-Gd distribution (orange) is broader than the pre-Gd one (blue), most likely due to susceptibility gradients produced by gadolinium.

No anterior-posterior trend from genu to splenium was discernible (see. **Supplementary Figure 1** for an example); furthermore in this work we focused exclusively on the genu.

### 3.4. WMTI-Watson model

Pre-Gd, the two branches of the WMTI-Watson model differed in the same ways as previously reported: one solution (Branch 1) was associated with $D_{a,\parallel} > D_{e,\parallel}$; the other solution (Branch 2) was characterized by $D_{a,\parallel} < D_{e,\parallel}$, and lower intra-axonal water fraction and dispersion (i.e. higher $c_2$) than Branch 1 (**Figure 6**). It should be noted only solutions within physically acceptable boundaries were retained, i.e. $f \in [0, 1], D_{a,\parallel}, D_{e,\parallel}, D_{e,\perp} \in [0, 4]\ \mu m^2/ms, c_2 \in \left[\frac{1}{3}, 1\right]$. Since $D_{a,\parallel}$ estimates were close to 3 $\mu m^2/ms$ in Solution 1, the upper bound on diffusivities was chosen to be 4 $\mu m^2/ms$ to avoid a truncation bias. The percentage of voxels in the genu that exhibited a solution within the defined bounds was (65±17)% for Branch 1 and (80±10)% for Branch 2. The gadolinium perfusion translated into a significant increase in intra-axonal water fraction for both branches: Branch 1: median $f^{pre/post}$ = 0.43/0.52, $p$ = 0.004; Branch 2: median $f^{pre/post}$ = 0.35/0.43, $p$ = 0.027, which was consistent with the intended effect of the procedure. Branch 2 also displayed a significant increase in $D_{a,\parallel}$ post-Gd ($p$ = 0.048).

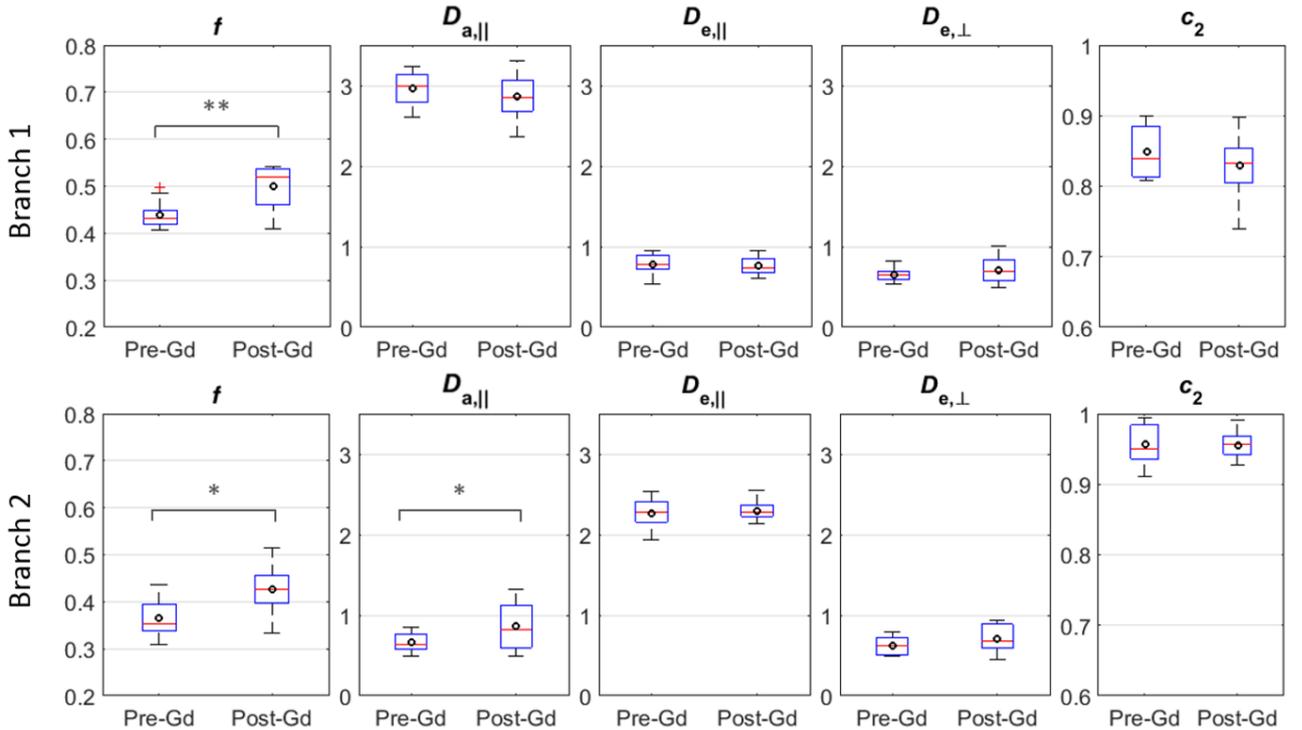

**Figure 6.** WMTI-Watson metrics in the genu, before and after gadolinium perfusion. The model showed increased intra-axonal water fraction post-Gd, for both sets of solutions. Branch 2 (bottom row) was associated with an increased intra-axonal diffusivity also. *Red line: median; box edges: 25th and 75th percentiles; whiskers: extreme datapoints; red cross: outliers; black circle: mean. \*: p < 0.05, \*\*: p < 0.01*. $f$ = intra-axonal water fraction; $D_{a,\parallel}$ = intra-axonal diffusivity $D_{e,\parallel}$ / $D_{e,\perp}$ = extra-axonal axial/radial diffusivity; and $c_2 \equiv \langle (\cos \psi)^2 \rangle$. All diffusivities in $\mu m^2/ms$.

## 3.5. Simulations

The median values of model parameters estimated pre-Gd were used as ground truths for pre-Gd signal simulation. Branch 1 ground truth was thus: $f = 0.43$, $D_{a,\|} = 3.0$, $D_{e,\|} = 0.78$, $D_{e,\perp} = 0.65$ and $c_2 = 0.84$. Branch 2 ground truth was: $f = 0.35$, $D_{a,\|} = 0.63$, $D_{e,\|} = 2.5$, $D_{e,\perp} = 0.62$ and $c_2 = 0.95$. The post-Gd change was simulated as an increase in $f$ from 0.43 to 0.52 (Branch 1), or from 0.35 to 0.43 (Branch 2), with all other model parameters unchanged. The SNR measured in the genu at $b = 0$ was $16 \pm 2$ pre-Gd and $17 \pm 4$ post-Gd, and was boosted to 37 by the denoising procedure, which was the value used in the simulations.

The simulations of both branches reproduced experimental estimates well (see **Figure 7** vs **Figure 6**). However, the simulations predicted parameter changes associated with a post-Gd increase in intra-axonal water fraction only, while the data were the result of all gadolinium-related effects (including susceptibility gradients).

Regarding tensor metrics, simulations also confirmed that an increase in intra-axonal fraction translated into an increase in radial kurtosis (**Supplementary Figure 3**). However, the resulting increase in AD (in the case of Branch 1) and decrease in AD (in the case of Branch 2) were larger than the noise and should have been detectable experimentally. Aside from biological variability, which the simulations do not account for, these results further point to the fact that gadolinium-induced susceptibility gradients have an important impact on the diffusion tensor estimates, and should be included in the interpretation of the results.

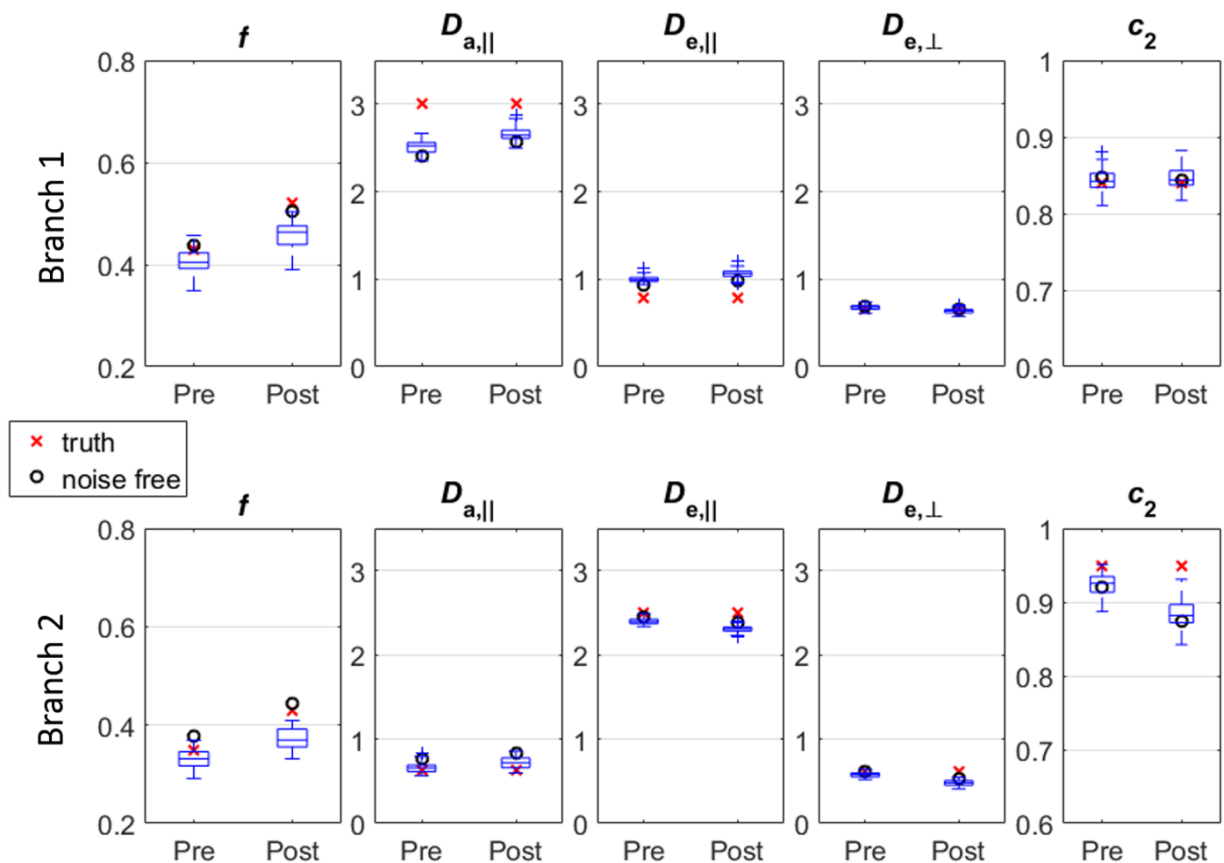

**Figure 7**. Simulations of WMTI-Watson model parameter estimates, assuming the post-Gd condition translated into an increased intra-axonal water fraction *f*. The noise free estimates (black circles) highlight the potential intrinsic bias in parameter estimation relative to the known ground truth (red crosses). The boxplots illustrate 50 estimates obtained using 1000 simulations with SNR = 37 and further averaging over 20 measurements to mimic ROI averaging. Simulations of both branches reproduced experimental data well (see **Figure 6**).

For Branch 1, $D_{a,\|}$ was systematically underestimated and $D_{e,\|}$ overestimated (**Figure 7**, top row, noise free values), with excellent accuracy for all other parameters. The $D_{a,\|}$ underestimation was more pronounced for lower intra-axonal fraction (Pre-Gd configuration). Noise further led to an underestimation of $f$, and a slight overestimation of $D_{a,\|}$ and $D_{e,\|}$ (relative to the noise free value). The precision for all parameters was very good ($\leq 6\%$). The two parameters where the post-Gd trends differed between data and simulation were $D_{a,\|}$ and $c_2$. Simulations predicted an increase in $D_{a,\|}$ (related to improved accuracy when the intra-axonal fraction was larger) while the experimental $D_{a,\|}$ was mildly reduced post-Gd. This suggests that gadolinium background gradients caused an underestimation of $D_{a,\|}$. The experimental orientation dispersion appeared somewhat increased (i.e. lower $c_2$) post-Gd, though the change was non-significant; this trend was not predicted by an increase in $f$ in the simulations. Thus gadolinium background gradients could contribute to an apparent higher intra-voxel orientation dispersion, similarly to the broader distribution of diffusion tensor main orientations relative to $\mathbf{B}_0$.

For Branch 2, $c_2$ was systematically underestimated and $D_{a,\|}$ overestimated (**Figure 7**, bottom row, noise free values). The $c_2$ underestimation was more pronounced for a higher intra-axonal fraction $f$ (post-Gd configuration). Noise further produced an underestimation of $f$ and slight underestimation of $D_{a,\|}$ (balancing the bias and making the noisy $D_{a,\|}$ estimates more accurate than the noise free ones). The precision for all parameters was $\leq 9\%$. The two parameters where the post-Gd trends differed between data and simulation were, as for Branch 1, $D_{a,\|}$ and $c_2$. An experimental increase in $D_{a,\|}$ was measured, which suggested that gadolinium background gradients would cause on overestimation of $D_{a,\|}$. No change in $c_2$ was measured while simulations predicted a decrease; thus gadolinium background gradients would contribute to increase $c_2$ (i.e. lower the apparent orientation dispersion).

## 4. Discussion

This study used Gd accumulation in the extracellular space of corpus callosum to vary the intra-axonal contribution to the total signal and evaluate the ensuing impact on a large number of diffusion parameters: main diffusion and kurtosis tensor metrics, but also compartment-specific parameters of a two-compartment model of diffusion. The goal of the study was to determine, using this gadolinium challenge, which of the two possible solutions of the two-compartment model (referred to as Branch 1: $D_{a,\|} > D_{e,\|}$ and Branch 2: $D_{a,\|} < D_{e,\|}$) is the biologically-relevant one. Results support the hypothesis that $D_{a,\|} > D_{e,\|}$, as detailed below.

First, the dramatic shortening in $T_2$ and $T_1$ relaxation times confirmed that gadolinium accumulated in the extracellular space of the genu. The measured $T_1$ was significantly longer for TE = 30 ms vs TE = 2.8 ms, both pre- and post-Gd. In the pre-Gd experiment, this can potentially be attributed to myelin water signal present at TE = 2.8 ms and suppressed at TE = 30 ms. In the post-Gd experiment, the variability was much larger – depending on the efficiency of each individual perfusion – but we hypothesize the short-TE measurement of $T_1$ still had contributions from all compartments, while at long TE the predominant contribution was from intracellular $T_1$. Silva et al. reported a similar result for post-Gd $T_1$ measurements [19]. The post-Gd long-TE $T_1$ estimate – which presumably reflects intracellular $T_1$ – was significantly shorter than the pre-Gd estimate. While it has been argued that intracellular $T_1$ could be significantly shorter than extracellular $T_1$ [19], it is also possible that water exchange between the two compartments, combined with inversion times of up to 6 seconds in the post-Gd experiments, precluded compartment selectivity in this measurement, as recently discussed [34].

The significant increase in radial kurtosis post-Gd is consistent with a selective attenuation of the extracellular signal fraction, as shown in simulations (**Supplementary Figure 3**). However, susceptibility gradients have also been reported to produce an overestimation of radial kurtosis in fibers perpendicular to $\mathbf{B}_0$ (such as corpus callosum) relative to fibers parallel to $\mathbf{B}_0$ [35], whereby the experimental increase in RK is likely a combination of both effects. This is further supported by the strong correlation between $T_2$ and RK.

The WMTI-Watson two-compartment model of diffusion revealed an increase in intra-axonal water fraction (as aimed by the gadolinium infusion) for both branches. However, we argue that Branch 1 ($D_{a,\parallel} > D_{e,\parallel}$) is the biologically relevant one. This conclusion is based on both the compatibility of pre-Gd estimates with previous measurements of rat corpus callosum, and the evolution of model parameters following gadolinium infusion compared to simulations and expected effects of susceptibility gradients on intra-axonal diffusivity.

Regarding pre-Gd values, the orientation dispersion estimated in Branch 1 ($c_2 = 0.84$, i.e. 24°) is closer to previous estimates of 34° for the dispersion in the corpus callosum of the rat [36] than the estimation provided by Branch 2 ($c_2 = 0.95$, i.e. 13°). Furthermore, simulations showed that the $c_2$ estimate was both accurate and precise in Branch 1, while $c_2$ was underestimated in Branch 2. This means that the experimental $c_2$ estimate in Branch 2, which was already rather on the high end, could potentially be an underestimation: thus Branch 2 would be associated with nearly perfectly aligned axons, which is unrealistic. It should however also be noted that, for Branch 1, the $D_{a,\parallel}$ estimate was rather high (3.0 µm²/ms) and simulations further showed $D_{a,\parallel}$ to be underestimated in Branch 1, bringing the "true" $D_{a,\parallel}$ to values higher than 3 µm²/ms. High intra-axonal diffusivity estimates could be caused by an unaccounted CSF compartment and an optimal body temperature of 38°C in the rat. Most recent estimates using alternative methods agree on $D_{a,\parallel} \approx 2.3 - 2.5$ µm²/ms, as will be discussed in more detail later on [21, 22, 37, 38].

The experimental change in model parameters following gadolinium infusion also favors the $D_{a,\parallel} > D_{e,\parallel}$ scenario. In Branch 1, a significant increase in $f$ is measured post-Gd ($p < 0.01$) while all other parameters are only slightly altered (below statistical significance). Simulations revealed that an increased intra-axonal water fraction should lead to an increase in $D_{a,\parallel}$ for this particular ground truth and acquisition protocol. Experimentally, no significant change in $D_{a,\parallel}$ was measured, with a slight trend towards reduced values. The major difference between experiment and simulations lies in the impact of Gd-induced susceptibility gradients, which are not accounted for in simulations and are expected to underestimate the diffusivity, as will be discussed shortly. Thus the increase in $D_{a,\parallel}$ expected from model bias was balanced by a post-Gd underestimation of $D_{a,\parallel}$ due to susceptibility gradients. Similarly for orientation dispersion, simulations predicted an increased intra-axonal fraction would not affect the $c_2$ estimate, which was highly accurate and precise. Experimentally, no significant change in $c_2$ was measured, with a slight trend towards reduced values (i.e. higher dispersion).

This trend could also be consistent with susceptibility gradients impacting orientation estimates at the intra-voxel level, in a similar way to the estimate at the whole-voxel level (see **Figure 5** for estimated angles between diffusion tensor and $\mathbf{B}_0$ field) [33]. Moving on to Branch 2, significant but weaker ($p < 0.05$) increases in both $f$ and $D_{a,\parallel}$ were measured experimentally. Simulations of increased intra-axonal fraction predicted no change in $D_{a,\parallel}$; thus the increase measured experimentally could be attributed to susceptibility gradients causing an overestimation of $D_{a,\parallel}$. This is opposite to the expected effect of such gradients on diffusivities. Furthermore, simulations predicted that an increase in intra-axonal fraction would cause a decrease in the $c_2$ estimation (bias), while no change in $c_2$ was measured experimentally. Thus, background gradients would have caused a lower apparent orientation dispersion (higher $c_2$), which is also contrary to expected effects.

Coming back on the influence of background gradients on diffusivity estimates, [39] have shown that, while individual isochromats may display overestimated or underestimated diffusivity depending on whether the background gradient is parallel or antiparallel to the diffusion-sensitizing gradients, isochromats with reduced diffusivity have a larger signal than those with increased diffusivity and thus more weighting in the overall signal. In the context of background gradients following a Gaussian-like distribution with zero mean, this results in an underestimation of the overall diffusion coefficient. Water inside the axons is expected to experience gradients from all directions – at least in the radial plane, and due to dispersion likely also with some axial component – and thus see its apparent diffusivity reduced. The underestimation of the diffusion coefficient in the presence of background gradients has been reported in other studies as well: magnetic field gradients induced by microvasculature on the diffusion measurement of tissue (intra/extra cellular) water, for example [40]. Fahrrer and colleagues recently studied "parallel fiber" phantoms of Dyneema fibers bathing in an aqueous solution with variable concentrations of magnesium chloride (to vary the susceptibility difference between the Dyneema and the surrounding medium). When the medium was plain water (which made for a large susceptibility difference with the Dyneema), both the axial and radial diffusivities were gradually and substantially underestimated with varying fiber orientations (from 0 to 90°) relative to the main field. The authors also put forward imperfect fiber alignment resulting in axial components of the background gradients as an explanation for the underestimation of AD. Therefore, in the context of the experiments presented in this work, gadolinium-based susceptibility gradients are expected to produce an underestimation of the intra-axonal diffusivity in the post-Gd experiments. This effect, which constitutes the main difference between our experiments and simulations, can only explain trends observed for Branch 1, and not Branch 2.

Background gradients introduced by gadolinium are thus a confounding effect for the interpretation of results. While their qualitative impact on the data can be predicted (i.e. an underestimation of diffusivity), the current work does not provide quantification for this effect. One way to reduce the impact of these gradients on the diffusion coefficient estimation would be to use bipolar gradients for diffusion encoding [41, 42]. However, the longer TE associated with this type of sequence would pose sensitivity problems at 14 T: the TE in this study (48 ms) – while requiring a four-shot segmented EPI read-out – was already long compared to the estimated $T_2$ of 25 ms pre-Gd and 14 ms post-Gd. Simulations accounting for the effect of these gradients would rely on several *ad hoc* assumptions which can dramatically impact their reliability. In particular, the effective concentration of gadolinium in the extra-cellular space is not known, and its estimation relies on the native $T_2$ and post-Gd $T_2$ of the extra-cellular compartment, as well as on the relaxivity in brain tissue of gadobutrol at 14 Tesla, all of which would be somewhat speculative.

We note that the increase in intra-axonal water fraction achieved with the gadolinium infusion – as estimated from the WMTI-Watson model – was only moderate. While water exchange across the axonal membrane would attenuate any difference between pre- and post-Gd measurements (by effectively restoring the signal fraction between intra- and extra-axonal compartments), permeability is expected to be negligible for myelinated axons over a diffusion time of 20 ms [43]. The low permeability assumption is also supported by the significant increase in RK post-Gd. On the other hand, $T_2$ reduction in the intra-axonal space due – again – to susceptibility gradients is expected to occur and result in a less pronounced intra-axonal water selection. Furthermore, in the absence of exchange, $T_1$ is expected to be shortened in the extra-cellular space only, which would also work against extra-cellular signal suppression by increasing the steady-state signal available in that compartment (TR = 2 s, native $T_1$ = 1.9 s, post-Gd $T_1$ = 0.3 s).

In addition to dispersion, axonal undulation is also known to impact diffusion measurements [44]. However, undulation is in fact predominant in extracranial white matter – to allow for mechanical stretching and compression – and the brain white matter tracts only display fascicular undulation, with wavelengths at least on the order of the voxel size in this study (e.g. 800 μm slice thickness) [45]. On these length scales, modeling undulation is equivalent to modeling intra-voxel orientation dispersion.

The results presented here are in agreement with several recent reports that used different approaches to address the same question regarding compartment axial diffusivities. Using isotropic diffusion weighting, it has been shown that isotropic kurtosis was negligible in most brain regions, including white matter tracts, whereby the compartment traces were similar [21, 24]. Given that $\mathrm{Tr}\,\widehat{D}_a = D_{a,\parallel}$ while $\mathrm{Tr}\,\widehat{D}_e = D_{e,\parallel} + 2D_{e,\perp}$, it follows that $D_{a,\parallel} > D_{e,\parallel}$. Selective suppression of extra-axonal water can in principle be achieved by exploiting $D_{a,\perp}=0$ and applying a very strong gradient orthogonally to the main bundle direction. A double diffusion encoding (DDE) sequence, using suppression along one orthogonal direction, was used in the rat spinal cord, and results suggested $D_{a,\parallel} \approx D_{e,\parallel}$ [23]. A planar filter was recently used to perform a more efficient suppression along all orthogonal directions, yielding an axonal diffusivity estimate $D_{a,\parallel} = 2.0$ μm$^2$/ms in the human white matter, in the infinite time limit [22]. Intra-axonal water selectivity achieved by ultra-high diffusion attenuation (up to 10 ms/μm$^2$) also led to an estimated interval of [1.9, 2.2] μm$^2$/ms for $D_{a,\parallel}$ [37]. A different approach has been to examine the time-dependence of compartment-specific diffusivities in a Watson-WMTI model, which has also shown that functional forms are physically acceptable only for the set of solutions corresponding to $D_{a,\parallel} > D_{e,\parallel}$ [20]. The latter work has been performed in fixed rat spinal cord, which suggests the inequality between compartment diffusivities holds both *in vivo* and *ex vivo*. Most recently, the fiber ball white matter modeling method has also output $D_{a,\parallel}$ values of 2.2 – 2.5 μm$^2$/ms [38].

The current study further showed that the trace of the diffusion tensor (or, equivalently, the mean diffusivity) in the corpus callosum did not change after gadolinium perfusion, which is similar to findings in gray matter using gadolinium-based [19] and fluorine-based [18] contrast agents. The similarity in compartment traces has been shown to be a distinctive feature of most brain regions with the exception of the thalamus [24], and can prove very useful for constraining model fitting, as already implemented by [46].

We therefore conclude that it is possible to constrain the two-compartment model of diffusion in white matter to solutions characterized by $D_{a,\parallel} \geq D_{e,\parallel}$ or similarity of compartment traces. However, validation of parameter values in various pathologies remains to be performed [12], as they could differ substantially from the healthy brain.

## 5. Conclusion

In this work, we used an intracerebroventricular injection of a gadolinium-based contrast agent to attenuate the extracellular signal in the rat brain, and compared diffusion, kurtosis, and WMTI-Watson model metrics in the genu of the corpus callosum before and after gadolinium infusion. The significant increase in radial kurtosis post-Gd suggested the relative fraction of extracellular water signal was indeed decreased by the procedure. This was further supported by a significant increase in intra-axonal water fraction as estimated from the two-compartment model, for both branches. However, pre-Gd estimates of axon dispersion in Branch 1 agreed better with literature than those of Branch 2. Furthermore, comparison of post-Gd changes in diffusivity and dispersion between data and simulations further supported Branch 1 as the biologically plausible solution, i.e. $D_{a,\parallel} > D_{e,\parallel}$. This result is fully consistent with other recent measurements of compartment axial diffusivities that used entirely different approaches.

## Acknowledgments

The authors thank Dmitry Novikov and Jelle Veraart for discussions on denoising and tensor fitting. This work was supported by the Centre d'Imagerie Bio-Médicale (CIBM) of the University of Lausanne (UNIL), the Swiss Federal Institute of Technology Lausanne (EPFL), the University of Geneva (UniGe), the Centre Hospitalier Universitaire Vaudois (CHUV), the Hôpitaux Universitaires de Genève (HUG) and the Leenaards and the Jeantet Foundations.

# References

1. Basser, P.J., J. Mattiello, and D. LeBihan, *Estimation of the effective self-diffusion tensor from the NMR spin echo.* J Magn Reson B, 1994. **103**(3): p. 247-54.
2. Jensen, J.H., et al., *Diffusional kurtosis imaging: the quantification of non-gaussian water diffusion by means of magnetic resonance imaging.* Magn Reson Med, 2005. **53**(6): p. 1432-40.
3. Stanisz, G.J., et al., *An analytical model of restricted diffusion in bovine optic nerve.* Magn Reson Med, 1997. **37**(1): p. 103-11.
4. Behrens, T.E., et al., *Characterization and propagation of uncertainty in diffusion-weighted MR imaging.* Magn Reson Med, 2003. **50**(5): p. 1077-88.
5. Jespersen, S.N., et al., *Modeling dendrite density from magnetic resonance diffusion measurements.* Neuroimage, 2007. **34**(4): p. 1473-86.
6. Zhang, H., et al., *NODDI: practical in vivo neurite orientation dispersion and density imaging of the human brain.* Neuroimage, 2012. **61**(4): p. 1000-16.
7. Fieremans, E., J.H. Jensen, and J.A. Helpern, *White matter characterization with diffusional kurtosis imaging.* Neuroimage, 2011. **58**(1): p. 177-88.
8. Wang, Y., et al., *Quantification of increased cellularity during inflammatory demyelination.* Brain, 2011. **134**(Pt 12): p. 3590-601.
9. Scherrer, B., et al., *Characterizing brain tissue by assessment of the distribution of anisotropic microstructural environments in diffusion-compartment imaging (DIAMOND).* Magn Reson Med, 2016. **76**(3): p. 963-77.
10. Panagiotaki, E., et al., *Compartment models of the diffusion MR signal in brain white matter: a taxonomy and comparison.* Neuroimage, 2012. **59**(3): p. 2241-54.
11. Ferizi, U., et al., *A ranking of diffusion MRI compartment models with in vivo human brain data.* Magn Reson Med, 2014. **72**(6): p. 1785-92.
12. Jelescu, I.O. and M.D. Budde, *Design and Validation of Diffusion MRI Models of White Matter.* Frontiers in Physics, 2017. **5**(61).
13. Novikov, D.S., et al., *Quantifying brain microstructure with diffusion MRI: Theory and parameter estimation.* ArXiv e-prints, 2016: p. arXiv:1612.02059v2 [physics.bio-ph].
14. Novikov, D.S., V.G. Kiselev, and S.N. Jespersen, *On modeling.* Magn Reson Med, 2018. **79**(6): p. 3172-3193.
15. Jelescu, I.O., et al., *Degeneracy in model parameter estimation for multi-compartmental diffusion in neuronal tissue.* NMR Biomed, 2016. **29**(1): p. 33-47.
16. Novikov, D.S., et al., *Rotationally-invariant mapping of scalar and orientational metrics of neuronal microstructure with diffusion MRI.* NeuroImage, 2018. **174**: p. 518-538.
17. Goodman, J.A., J.J. Ackerman, and J.J. Neil, *Cs+ ADC in rat brain decreases markedly at death.* Magn Reson Med, 2008. **59**(1): p. 65-72.
18. Duong, T.Q., et al., *Evaluation of extra- and intracellular apparent diffusion in normal and globally ischemic rat brain via 19F NMR.* Magn Reson Med, 1998. **40**(1): p. 1-13.
19. Silva, M.D., et al., *Separating changes in the intra- and extracellular water apparent diffusion coefficient following focal cerebral ischemia in the rat brain.* Magn Reson Med, 2002. **48**(5): p. 826-37.
20. Jespersen, S.N., et al., *Diffusion time dependence of microstructural parameters in fixed spinal cord.* Neuroimage, 2017.
21. Dhital, B., et al., *The absence of restricted water pool in brain white matter.* Neuroimage, 2017.
22. Dhital, B., et al., *Intra-axonal diffusivity in brain white matter.* arXiv:1712.04565 [physics.bio-ph], 2017.
23. Skinner, N.P., et al., *Rapid in vivo detection of rat spinal cord injury with double-diffusion-encoded magnetic resonance spectroscopy.* Magn Reson Med, 2017. **77**(4): p. 1639-1649.
24. Szczepankiewicz, F., et al., *Quantification of microscopic diffusion anisotropy disentangles effects of orientation dispersion from microstructure: applications in healthy *volunteers and in brain tumors.* Neuroimage, 2015. **104**: p. 241-52.
25. Veraart, J., E. Fieremans, and D.S. Novikov, *Universal power-law scaling of water diffusion in human brain defines what we see with MRI.* ArXiv e-prints, 2016: p. arXiv:1609.09145 [physics.bio-ph].
26. Gruetter, R., *Automatic, localized in vivo adjustment of all first- and second-order shim coils.* Magn Reson Med, 1993. **29**(6): p. 804-11.
27. Gruetter, R. and I. Tkac, *Field mapping without reference scan using asymmetric echo-planar techniques.* Magn Reson Med, 2000. **43**(2): p. 319-23.
28. van de Looij, Y., et al., *Diffusion tensor echo planar imaging using surface coil transceiver with a semiadiabatic RF pulse sequence at 14.1T.* Magn Reson Med, 2011. **65**(3): p. 732-7.
29. Vanhamme, L., A. van den Boogaart, and S. Van Huffel, *Improved method for accurate and efficient quantification of MRS data with use of prior knowledge.* J Magn Reson, 1997. **129**(1): p. 35-43.
30. Veraart, J., et al., *Denoising of diffusion MRI using random matrix theory.* Neuroimage, 2016. **142**: p. 394-406.
31. Koay, C.G., E. Özarslan, and P.J. Basser, *A signal transformational framework for breaking the noise floor and its applications in MRI.* Journal of Magnetic Resonance, 2009. **197**(2): p. 108-119.
32. Veraart, J., et al., *Weighted linear least squares estimation of diffusion MRI parameters: strengths, limitations, and pitfalls.* Neuroimage, 2013. **81**: p. 335-46.
33. Bammer, R., et al., *Analysis and generalized correction of the effect of spatial gradient field distortions in diffusion-weighted imaging.* Magnetic Resonance in Medicine, 2003. **50**(3): p. 560-569.
34. Shazeeb, M.S. and C.H. Sotak, *Limitations in biexponential fitting of NMR inversion-recovery curves.* J Magn Reson, 2017. **276**: p. 14-21.
35. Palombo, M., et al., *New insight into the contrast in diffusional kurtosis images: does it depend on magnetic susceptibility?* Magn Reson Med, 2015. **73**(5): p. 2015-24.
36. Leergaard, T.B., et al., *Quantitative histological validation of diffusion MRI fiber orientation distributions in the rat brain.* PLoS One, 2010. **5**(1): p. e8595.
37. Veraart, J., D.S. Novikov, and E. Fieremans, *TE dependent Diffusion Imaging (TEdDI) distinguishes between compartmental T2 relaxation times.* Neuroimage, 2017.
38. McKinnon, E.T., J.A. Helpern, and J.H. Jensen, *Modeling white matter microstructure with fiber ball imaging.* NeuroImage, 2018. **176**: p. 11-21.
39. Zhong, J., R.P. Kennan, and J.C. Gore, *Effects of susceptibility variations on NMR measurements of diffusion.* Journal of Magnetic Resonance, 1991. **95**: p. 267-280.
40. Kiselev, V.G., *Effect of magnetic field gradients induced by microvasculature on NMR measurements of molecular self-diffusion in biological tissues.* Journal of Magnetic Resonance, 2004. **170**(2): p. 228-235.
41. Does, M.D., J. Zhong, and J.C. Gore, *In vivo measurement of ADC change due to intravascular susceptibility variation.* Magnetic Resonance in Medicine, 1999. **41**(2): p. 236-240.
42. Hong, X. and W. Thomas Dixon, *Measuring diffusion in inhomogeneous systems in imaging mode using antisymmetric sensitizing gradients.* Journal of Magnetic Resonance (1969), 1992. **99**(3): p. 561-570.
43. Nilsson, M., et al., *Noninvasive mapping of water diffusional exchange in the human brain using filter-exchange imaging.* Magn Reson Med, 2013. **69**(6): p. 1573-81.
44. Nilsson, M., et al., *The importance of axonal undulation in diffusion MR measurements: a Monte Carlo simulation study.* NMR Biomed, 2012. **25**(5): p. 795-805.
45. Gray, H., P. Williams, and L. Bannister, *Gray's Anatomy: the Anatomical Basis of Medicine and Surgery.* 1995, Edinburgh: Churchill Livingstone.
46. Reisert, M., et al., *Disentangling micro from mesostructure by diffusion MRI: A Bayesian approach.* Neuroimage, 2017. **147**: p. 964-975.

**Supplementary Figures**

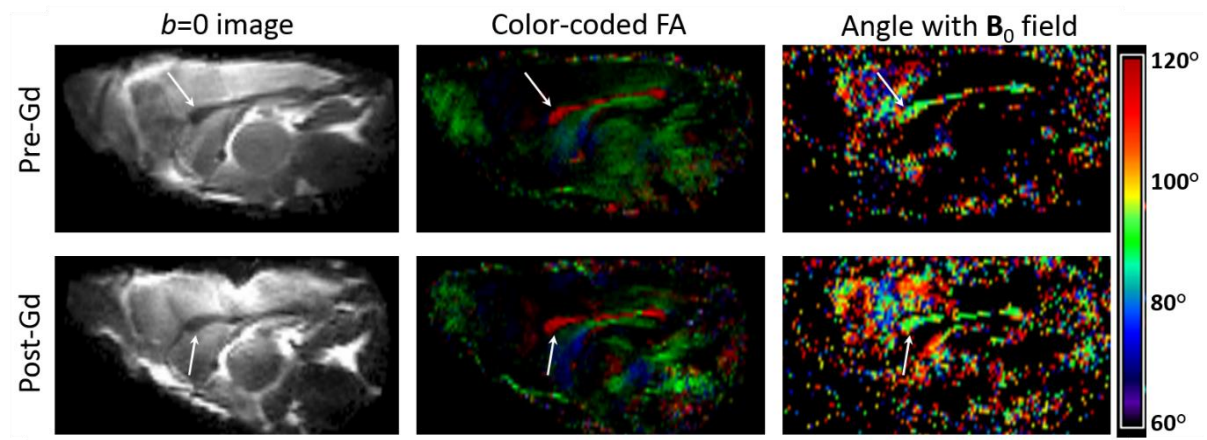

**Supplementary Figure 1**: Pre- and post-Gd $b = 0$ images of the mid-sagittal slice (left), matching color-coded FA maps (middle) (Red: L-R, Green: H-F, Blue: A-P), and angle between tensor principle direction and main magnetic field (right). The white arrows indicate the genu of the corpus callosum. Distortions due to the surgery and the accumulation of gadolinium can be seen on the post-Gd image, but the genu of the corpus callosum was not visibly affected. The angle between the diffusion tensor and the B0 field was around 90°, as expected.

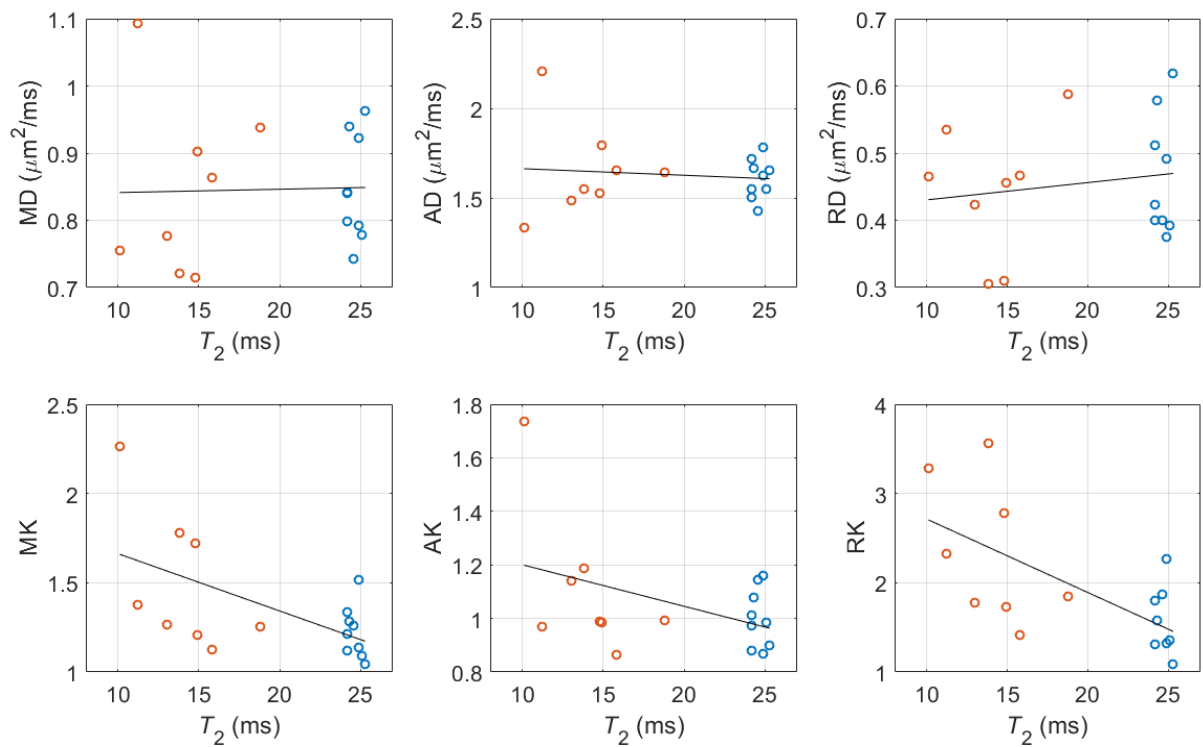

**Supplementary Figure 2**: Scatter plots of diffusion and kurtosis metrics vs $T_2$ in the genu. Blue: pre-Gd, red: post-Gd. $T_2$ is used as a proxy for gadolinium concentration in the post-Gd measurements. Radial kurtosis correlated significantly with $T_2$, (Pearson's $\rho$ = -0.66; $p <$ 0.01), but not axial kurtosis nor any diffusivity (black lines).

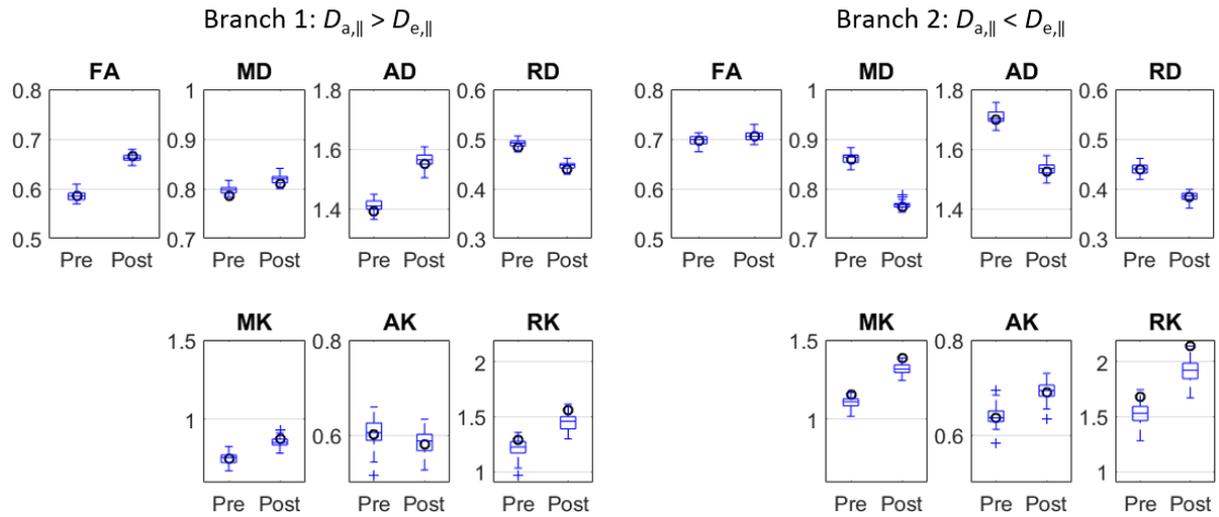

**Supplementary Figure 3**: Simulated diffusion and kurtosis tensor metrics assuming a WMTI-Watson model ground truth based on Branch 1 (left) or Branch 2 (right) (see Section 3.5 for ground truth values). In the case of Branch 1, an increase in the intra-axonal water fraction resulted in an increased FA, AD and RK, and a decreased RD. In the case of Branch 2, an increase in the intra-axonal water fraction resulted in a decrease in all diffusivities and increase in all kurtoses, and a decreased RD. For both branches though, only the increase in RK was significant experimentally. This result highlights the involvement of susceptibility gradients in the post-Gd measurement and the necessity to use compartment modeling to better understand their effect on intra-axonal water.

**Supplementary Data 1**:

The Watson distribution is characterized by a concentration parameter κ, from which one can directly derive $\langle(\cos\psi)^2\rangle \equiv c_2$:

$$F(x) = \frac{\sqrt{\pi}}{2} e^{-x^2} \text{erfi}(x) \tag{1}$$

$$c_2 = \frac{1}{2\sqrt{\kappa} F(\sqrt{\kappa})} - \frac{1}{2\kappa} \tag{2}$$

Perfectly aligned axons correspond to $\kappa = \infty$ and $c_2 = 1$, while isotropically-distributed axons correspond to $\kappa = 0$ and $c_2 = 1/3$.

The 2$^{nd}$ and 4$^{th}$ order spherical harmonics expansion coefficients of this axially-symmetric ODF can be expressed as [20]:

$$p_2 = \frac{3c_2 - 1}{2} \tag{3}$$

$$p_4 = c_2 \left(\frac{5}{8} - \frac{105}{16\kappa}\right) + \frac{35}{16\kappa} + \frac{3}{8} \tag{4}$$

The Legendre expansion coefficients of the diffusion and kurtosis tensors D and W can be directly related to the model parameters $f$, $D_{a,\parallel}$, $D_{e,\parallel}$, $D_{e,\perp}$ and κ [16, 20]:

$$D_0 = \frac{1}{3}\left(fD_{a,\parallel} + (1-f)(D_{e,\parallel} + 2D_{e,\perp})\right) \tag{5}$$

$$D_2 = \frac{2}{3}p_2\left(fD_{a,\parallel} + (1-f)(D_{e,\parallel} - D_{e,\perp})\right) \tag{6}$$

$$W_0 = 3\left\{\frac{1}{5D_0^2}\left[fD_{a,\parallel}^2 + (1-f)\left[5D_{e,\perp}^2 + (D_{e,\parallel} - D_{e,\perp})^2 + \frac{10}{3}D_{e,\perp}(D_{e,\parallel} - D_{e,\perp})\right] - D_2^2\right] - 1\right\} \tag{7}$$

$$W_2 = \left\{p_2\left[fD_{a,\parallel}^2 + (1-f)\left((D_{e,\parallel} - D_{e,\perp})^2 + \frac{7}{3}D_{e,\perp}(D_{e,\parallel} - D_{e,\perp})\right)\right] - \frac{1}{2}D_2(D_2 + 7D_0)\right\}\frac{12}{7D_0^2} \tag{8}$$

$$W_4 = \left\{p_4\left[fD_{a,\parallel}^2 + (1-f)(D_{e,\parallel} - D_{e,\perp})^2\right] - \frac{9D_2^2}{4}\right\}\frac{24}{35D_0^2} \tag{9}$$

Now relating the latter to main tensor metrics:

$$\text{AD} = D_0 + D_2 \tag{10}$$

$$\text{RD} = D_0 - \frac{D_2}{2} \tag{11}$$

$$\text{AK} = \left(\frac{D_0}{\text{AD}}\right)^2 \cdot W_\parallel = \left(\frac{D_0}{\text{AD}}\right)^2 \cdot (W_0 + W_2 + W_4) \tag{12}$$

$$\text{RK} = \left(\frac{D_0}{\text{RD}}\right)^2 \cdot W_\perp = \left(\frac{D_0}{\text{RD}}\right)^2 \cdot \left(W_0 - \frac{W_2}{2} + \frac{3}{8}W_4\right) \tag{13}$$

where AD = axial diffusivity, RD = radial diffusivity, AK = axial excess kurtosis, RK = radial excess kurtosis.